\documentclass[twocolumn]{jpsj3}
\usepackage{graphicx}
\usepackage{bm}
\usepackage{cite}
\usepackage{color}
\usepackage{amssymb}
\usepackage{amsmath}
\usepackage{revsymb}

\title{Metal-Insulator Transition of Dirac Fermions: Variational Cluster Study}
\author{Masaki Ebato, Tatsuya Kaneko, and Yukinori Ohta}
\inst{Department of Physics, Chiba University, Chiba 263-8522, Japan}
\date{\today}

\abst{
A comparative study is made on the metal-insulator transition of Dirac 
fermions in the honeycomb and $\pi$-flux Hubbard models at half filling 
by means of the variational cluster approximation and cluster dynamical 
impurity approximation.  Paying particular attention to the choice of 
the geometry of solver clusters and the inclusion of particle-bath sites, 
we show that the direct transition from the Dirac semimetallic state 
to the antiferromagnetic Mott insulator state occurs in these models, 
and therefore, the spin liquid phase is absent in the intermediate 
region, in agreement with recent quantum-Monte-Carlo--based 
calculations.  
}

\begin{document}
\maketitle

\section{Introduction}

The mechanism of the metal-insulator transition (MIT) has long been 
one of the central issues in strongly correlated electron systems.\cite{M90,IFT98}  
In particular, the MIT in correlated Dirac fermion systems has attracted 
much attention recently, a typical example of which is the honeycomb-lattice 
Hubbard model at half filling representing graphene.\cite{graphene}  
Because the honeycomb lattice is bipartite and free from frustration, 
the N\'eel antiferromagnetic (AF) Mott insulator (MI) state is realized 
in the strong coupling region.  However, unlike in the square-lattice Hubbard model, 
where perfect Fermi surface nesting is present, one expects that 
the AF order will not appear in the weak coupling region but rather 
the massless Dirac semimetallic (SM) state will be maintained until a 
critical interaction strength is reached.\cite{ST92,H06,MK09,J09} 

The MIT in the honeycomb lattice was studied by Meng \textit{et al.}\cite{MLWAM10} 
using the quantum Monte Carlo (QMC) method, whereby they claimed the 
presence of a quantum spin liquid (SL) state (or nonmagnetic MI state) in 
the intermediate region between the Dirac SM state and the antiferromagnetic 
Mott insulator (AFMI) state.  
Their study attracted much interest because it suggested the emergence 
of the SL state in systems without frustration in their spin degrees 
of freedom.  
However, subsequent studies based on the large-scale QMC method\cite{SOY12}, 
the pinning field approach using the QMC method,\cite{AH13} and analysis of the quantum 
criticality by finite-size scaling \cite{AH13,THAH14} have consistently 
suggested the direct transition from the SM state to the AFMI state, and 
therefore we now anticipate that the SL state is absent in this model.  
Similar debates have also been had for the $\pi$-flux Hubbard model, 
another Dirac fermion system, whereby the direct transition from the SM state
to the AFMI state is now anticipated.\cite{CS12,IAS14,THAH14}  

Quantum cluster methods have also been used to study the MIT in the 
honeycomb Hubbard model.\cite{WCTTL10,L11,LI12,YXL11,WRLL12,HL12,SO12,HS13,LW13,CBSKC14,LRTR14} 
In particular, cluster dynamical mean-field theory (CDMFT) and 
variational cluster approximation (VCA) calculations have shown that 
if the 6-site hexagonal ring is used as a solver cluster, the 
single-particle band gap opens even in the weak coupling region where 
the AF order is absent, thereby suggesting the presence of the SL 
state.\cite{YXL11,WRLL12,HL12,SO12} 
However, the opening of the band gap at the infinitesimal interaction 
strength was questioned,\cite{HL12,SO12,HS13} and 
moreover, from comparison with the results of the cluster dynamical 
impurity approximation (CDIA) and dynamical cluster approximation (DCA), 
the emergence of the nonmagnetic insulator phase predicted by the CDMFT 
and VCA was considered to be unrealistic.\cite{HS13,LW13,CBSKC14,LRTR14}  
So far, not much is known about the MIT in the $\pi$-flux Hubbard model 
studied by quantum cluster methods.  

In this paper, motivated by the above development in the field, we will 
make a comparative study on the MIT of correlated Dirac fermions in the 
honeycomb and $\pi$-flux Hubbard models at half filling by means of the VCA 
and CDIA.  
We will, in particular, point out that a suitable choice of the cluster geometry is essential in the quantum cluster calculations to suppress the opening of the band gap in the weak-coupling region and that the 
inclusion of particle-bath sites is important in discussing the 
order of the MIT as well as the transfer of spectral weight in the 
single-particle spectral function.  
We will thereby show that the direct transition from the Dirac SM state 
to the AFMI state occurs with increasing interaction strength in 
these models and that the SL phase is absent in their intermediate coupling 
region.  

\begin{figure}[!t]
\begin{center}
\includegraphics[width=0.95\columnwidth]{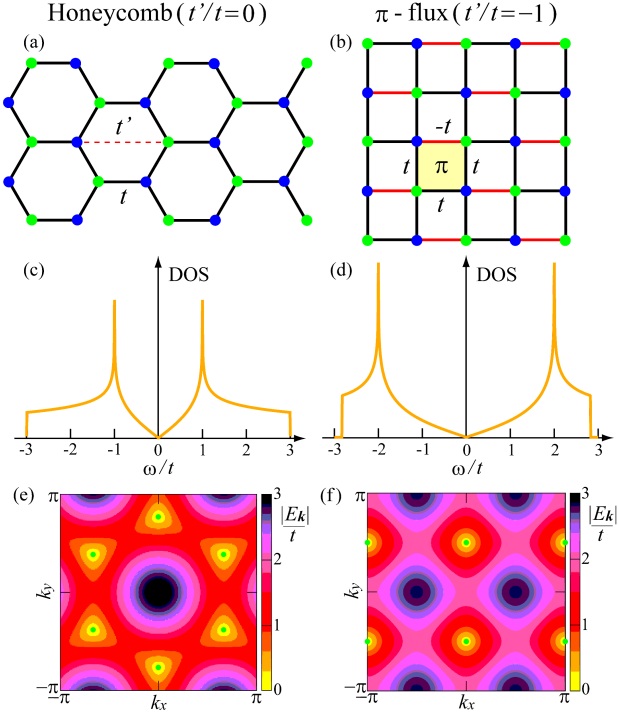}
\caption{(Color online) 
Schematic representations of the (a) honeycomb ($t'/t=0$) and 
(b) $\pi$-flux ($t'/t=-1$) lattices.  
The dashed line in (a) indicates the bonds with the hopping parameter 
$t'$ and the red lines in (b) indicate the bonds with the negative 
hopping parameter $t'=-t$.  Noninteracting DOSs [(c) and (d)] 
and contour plots of the band dispersions $E_{\bm k}$ [(e) and (f)] 
are also shown for the honeycomb (left panels) and $\pi$-flux 
(right panels) lattices.  The green dots in (e) and (f) indicate 
the Dirac points in ${\bm k}$-space.  
}\label{fig1}
\end{center}
\end{figure}

\section{Models and Methods}

The honeycomb and $\pi$-flux Hubbard models may be defined by the Hamiltonian 
\begin{equation}
\mathcal{H}= -\sum_{ i,j,\sigma}  t_{ij}c^{\dag}_{i\sigma}c_{j\sigma} 
+U \sum_{i}n_{i\uparrow}n_{i\downarrow},
\end{equation}
where $c^{\dag}_{i\sigma}$ is the creation operator of a fermion (which 
will be referred to as an electron hereafter) with spin $\sigma$ at site 
$i$ and $n_{i\sigma}=c^{\dag}_{i\sigma}c_{i\sigma}$.  
$t_{ij}$ is the hopping amplitude: we define $t_{ij}=t$ for the 
nearest-neighbor bonds and $t_{i,j}=t'$ for the bonds connecting 
hexagons in the honeycomb lattice [see Fig.~\ref{fig1}(a)].  
$U$ is the on-site Coulomb repulsion.  
We assume the filling of one electron per site (half filling).  
Changing the value of $t'$, we can tune the system continuously from 
the honeycomb lattice at $t'=0$ to the $\pi$-flux lattice at $t'=-t$ 
[see Fig.~\ref{fig1}(b)].\cite{HFA06}  
At $U=0$, these systems at low energies are described in terms 
of the massless Dirac fermions; their densities of states (DOSs) 
and band dispersions are shown in Figs.~\ref{fig1}(c)-\ref{fig1}(f).  

We apply the VCA,\cite{PAD03,S08,P12} which is a quantum 
cluster method based on self-energy functional theory (SFT).\cite{P04_1,P04_2}  
In the VCA, we introduce disconnected finite-size clusters 
(that are solved exactly) as a reference system.  
By restricting the trial self-energy to the self-energy of the reference 
system $\Sigma'$, we can obtain the grand potential of the original 
system in the thermodynamic limit as
\begin{align}
\Omega=\Omega'+\mathrm{Tr}\: \mathrm{ln} ( G^{-1}_0-\Sigma' )^{-1} - \mathrm{Tr}\: \mathrm{ln}( G' ), \label{gp}
\end{align}
where $\Omega'$, $G'$, and $G_0$ are the grand potential, the 
Green's function of the reference system, and the noninteracting Green's 
function, respectively.  
The short-range correlations within the cluster of the reference system 
are taken into account exactly.  
The one-body parameters $\bm{t}'$ of the reference system are 
optimized according to the variational principle 
$\partial \Omega[\Sigma'(\bm{t}')]/\partial \bm{t}' = 0$.  
In the VCA, we can treat the spontaneous symmetry breaking by adding 
appropriate Weiss fields to the reference system.\cite{DAHAP04}  
We have to choose an exactly solvable reference system; here we 
apply an exact diagonalization method and solve the quantum many-body 
problem in the cluster of the reference system.  

We also use the CDIA,\cite{S12} which is an extended version 
of the VCA where particle-bath sites are added to the clusters to 
take into account the electron-number fluctuations in the correlation 
sites.  In the CDIA, we optimize the hybridization parameter between the 
bath and correlation sites $V$ and the on-site energy of the bath 
sites $\varepsilon$ based on SFT.\cite{BKSTP09,S12}  
Note that the CDIA is intrinsically equivalent to CDMFT with an 
exact-diagonalization solver.  

\begin{figure}[!t]
\begin{center}
\includegraphics[width=\columnwidth]{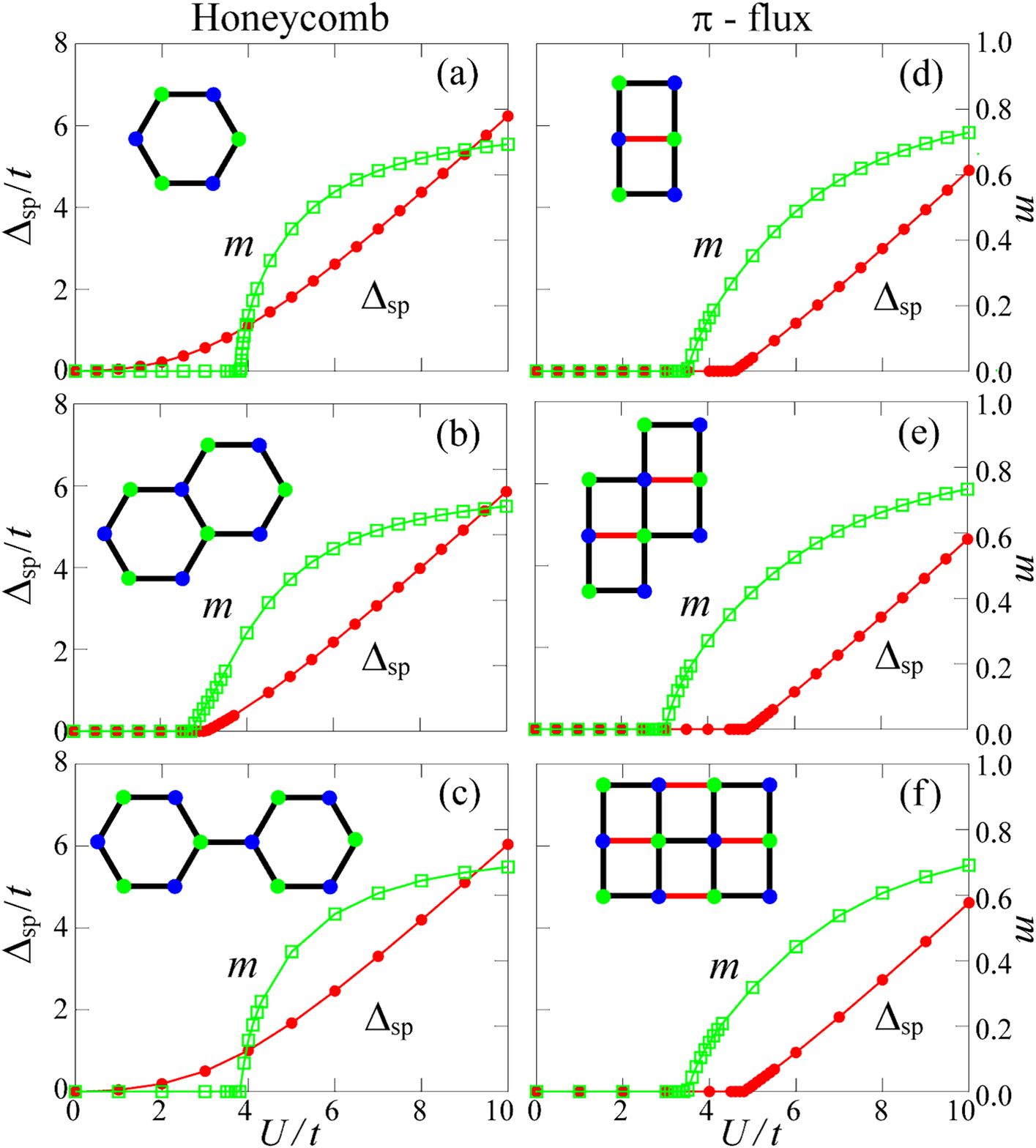}
\caption{(Color online) 
Single-particle gap $\Delta_{\mathrm{sp}}$ in the PM state and 
local magnetization $m$ in the AF state calculated by the VCA as 
functions of $U/t$.  
The results for the honeycomb Hubbard model are shown in (a), (b), 
and (c) and those for the $\pi$-flux Hubbard model are shown in 
(d), (e), and (f).  The geometry of the solver cluster used is 
illustrated in each panel.  
}\label{fig2}
\end{center}
\end{figure}

\begin{figure}[!t]
\begin{center}
\includegraphics[width=\columnwidth]{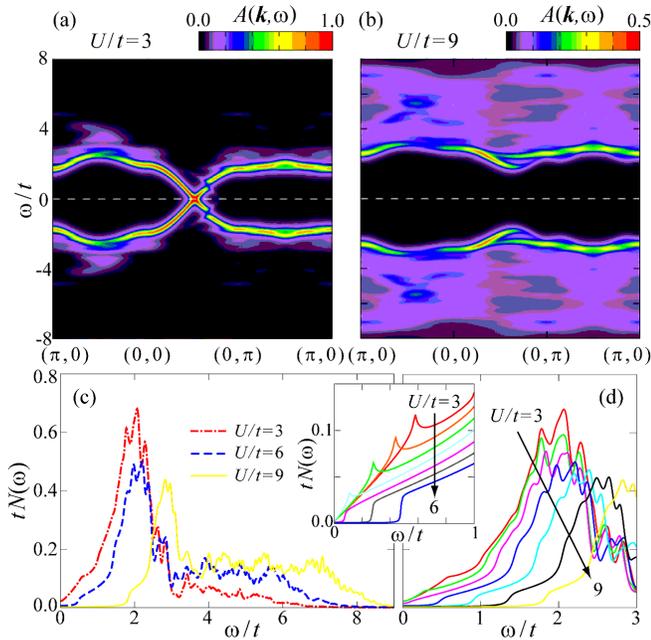}
\caption{(Color online) 
Single-particle spectral function $A({\bm k},\omega)$ [(a) and (b)] 
and DOS $N(\omega)$ [(c) and (d)] in the PM states of the $\pi$-flux 
Hubbard model calculated by the VCA with the 12-site cluster.  
The horizontal dashed line in (a) and (b) indicates the Fermi level. 
We applied the artificial Lorentzian broadening of the spectra of 
$\eta/t=0.15$ in (a) and (b) and $\eta/t=0.05$ in (c) and (d).  
The inset in the lower panels is an enlargement of the DOS near the Fermi level, 
assuming $\eta/t=0.005$.  
}\label{fig3}
\end{center}
\end{figure}

\section{Results and Discussion}

\subsection{Results of VCA}

First, let us consider the single-particle gap $\Delta_{\mathrm{sp}}$ 
in the paramagnetic (PM) state and the staggered magnetization $m$ in 
the AF state, which are calculated for the honeycomb and $\pi$-flux 
Hubbard models using the VCA.  We introduce the Weiss field associated with 
the two-sublattice N\'eel order and evaluate the local magnetization 
$m=\langle n_{i\uparrow} -n_{i\downarrow} \rangle$.  
The gap $\Delta_{\mathrm{sp}}$ is evaluated in the absence of the Weiss 
field as the jump of the chemical potential with respect to the number of 
electrons in the system.  Note that the band gap always opens 
when the AF order appears.  We use clusters of 6, 10, and 12 
sites as reference systems; the clusters used for the honeycomb and 
$\pi$-flux lattices are topologically equivalent but with different 
hopping parameters (see Fig.~\ref{fig2}).  

The results for the honeycomb Hubbard model are shown in Figs.~\ref{fig2}(a)-\ref{fig2}(c).  We find that the MIT is sensitive to the choice of the clusters, 
i.e., the results obtained using the clusters of 6 and 12 sites are 
qualitatively different from those in the case of 10 sites.  
The AF order appears at $U_{\mathrm{AF}}/t=3.8$ for the clusters of 
6 and 12 sites, the results of which are in good agreement with 
results of QMC simulations.\cite{SOY12,AH13}  
However, the gap $\Delta_{\mathrm{sp}}$ opens at infinitesimal $U$ values 
and the SM phase appears only at $U=0$, thus suggesting the presence of 
the PM insulator state at $0<U<U_{\mathrm{AF}}$.  
Recent studies, however, have claimed that this gap cannot be regarded as the 
true Mott gap,\cite{HS13,LW13} the details of which will be discussed 
below.  
For the cluster of 10 sites, on the other hand, the SM phase 
persists up to a large $U$ value and the transition to the AF phase 
occurs directly from the SM phase.  Here, the AF order appears at 
$U_{\mathrm{AF}}/t= 2.7$ and the gap $\Delta_{\mathrm{sp}}$ opens at 
$U_{\mathrm{PM}}/t=3.0$, qualitatively consistent 
with the results of recent QMC simulations, where the direct transition 
from the Dirac SM phase to the AFMI phase was predicted.\cite{SOY12,AH13} 

The results for the $\pi$-flux Hubbard model are shown in Figs.~\ref{fig2}(d)-\ref{fig2}(f).  We find that the results obtained using the clusters of 6, 
10, and 12 sites are qualitatively the same as each other, i.e., the 
SM phase persists up to a large $U$ value.  The AF order appears at 
$U_{\mathrm{AF}}/t=3.4$, 2.9, and 3.4 and the gap $\Delta_{\mathrm{sp}}$ 
opens at $U_{\mathrm{PM}}/t=4.5$, 4.9, and 4.8 for the clusters of 
6, 10, and 12 sites,  respectively.  
Therefore, the PM insulator state does not exist between the SM and AFMI 
phases, in accordance with the results of recent QMC simulations 
that show the direct transition from the SM phase to the AFMI phase.\cite{IAS14,THAH14}  
Note that the transition point $U_{\mathrm{AF}}$ of our VCA 
calculations is smaller than that of the QMC simulations, 
$U_{\mathrm{AF}}/t=5.25$--$5.5$,\cite{IAS14} which may be due to the 
anisotropy of the clusters used in our calculations; the agreement 
becomes good if we use an isotropic cluster of $4$ sites, which gives 
the value $U_{\mathrm{AF}}/t=5.0$.  

We also calculate the single-particle spectral function and DOS using 
cluster perturbation theory (CPT)\cite{SPP00} for the PM state of 
the $\pi$-flux Hubbard model.  The results are shown in Fig.~\ref{fig3}.  
We immediately find that the Dirac linear band dispersion is clearly visible 
near the Fermi level at $U<U_\mathrm{PM}$ [see Fig.~\ref{fig3}(a)], 
whereas the band gap opens at $U>U_\mathrm{PM}$ [see Fig.~\ref{fig3}(b)].  
The transfer of spectral weight occurring with increasing $U/t$ is seen 
in Figs.~\ref{fig3}(c) and \ref{fig3}(d), which is characteristic of the VCA and 
will be discussed below in Sect. 3.2 in comparison with the results of the CDIA.  

\begin{figure}[!t]
\begin{center}
\includegraphics[width=\columnwidth]{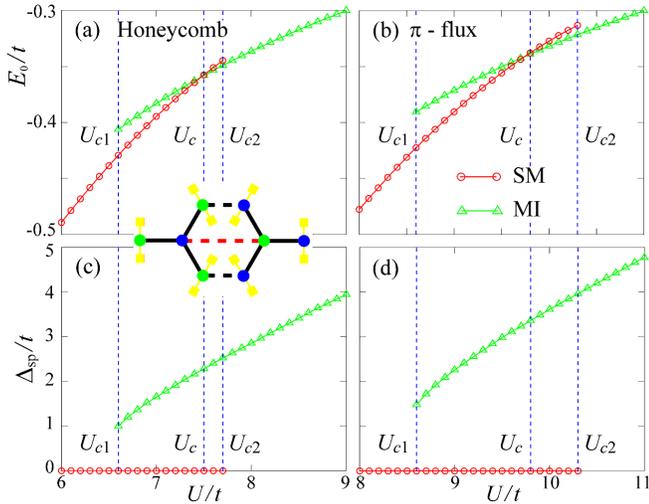}
\caption{(Color online)
Ground-state energies [(a) and (b)] and single-particle gaps [(c) and (d)] 
in the PM state of the honeycomb and $\pi$-flux Hubbard models calculated 
by the CDIA, where the 4-site 6-bath cluster shown in the left panels is used.  
}\label{fig4}
\end{center}
\end{figure}

\subsection{Results of CDIA}

Next, let us discuss the roles of bath sites in MIT using the CDIA.  
Following a previous study on the honeycomb Hubbard model,\cite{HS13} 
we examine the honeycomb and $\pi$-flux Hubbard models using the 
4-site 6-bath cluster.  The results are shown in Fig.~\ref{fig4}.  
We find that the grand potentials of the honeycomb and $\pi$-flux 
Hubbard models both have two stationary points around the transition 
point $U_c$.  
The SM solution exists at a small $U$, which vanishes at $U_{c2}$ with increasing $U$.  
The MI solution exists at a large $U$, which vanishes at $U_{c1}$ with decreasing $U$.  
The two solutions thus coexist in the region $U_{c1}\le U \le U_{c2}$, and 
the ground-state energies cross at $U_c$.  
We obtain the values $U_{c1}/t=6.6$, $U_{c2}/t=7.7$, and $U_{c}/t=7.5$ 
for the honeycomb Hubbard model and 
$U_{c1}/t=8.6$, $U_{c2}/t=10.3$, and $U_{c}/t=9.8$ 
for the $\pi$-flux Hubbard model.  
The calculated result for $\Delta_{\mathrm{sp}}$ [see Figs.~\ref{fig4}(c) and \ref{fig4}(d)] 
shows hysteresis between the SM ($\Delta_{\mathrm{sp}}=0$) and MI 
($\Delta_{\mathrm{sp}}>0$) solutions, which indicates that 
$\Delta_{\mathrm{sp}}$ jumps discontinuously at $U_c$.  
These results clearly indicate that the MIT is of the first-order 
(or discontinuous) in the CDIA, which is in contrast to the results of 
the VCA where the second-order (or continuous) transition is found 
(see Fig.~\ref{fig2}).  
The first-order MIT is thus expected in the actual honeycomb and 
$\pi$-flux Hubbard models in which the electron-number fluctuation 
is present.  

\begin{figure}[!t]
\begin{center}
\includegraphics[width=\columnwidth]{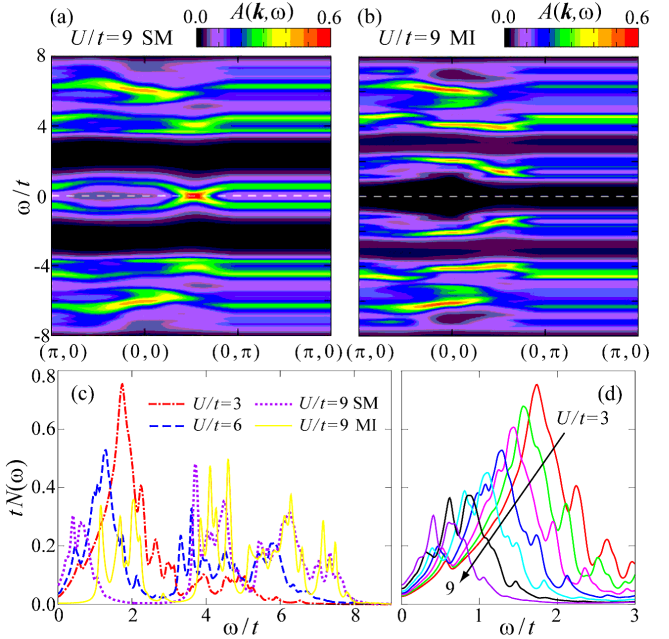}
\caption{(Color online) 
As in Fig.~\ref{fig3} but for the results of the CDIA with the 
4-site 6-bath cluster.  
In (a)-(c), the single-particle spectral functions and DOSs of 
both the SM and MI states are shown at $U/t=9$, while in (d), 
the DOS of only the SM state is shown.  
}\label{fig5}
\end{center}
\end{figure}

To further clarify the roles of bath sites in the MIT, we examine the $U$ 
dependence of the single-particle spectral function and the DOS calculated 
using CPT.   
The results for the $\pi$-flux Hubbard model obtained in the CDIA are 
shown in Fig.~\ref{fig5}, where the Dirac linear band dispersion is 
clearly visible in the vicinity of the Fermi level.  Note that 
the slope of the dispersion at the Dirac point becomes steeper for 
larger values of $U$.  

Comparing the DOS curves, we find that the results in the VCA (see Fig.~\ref{fig3}) 
are indeed significantly different from those in the CDIA (see Fig.~\ref{fig5}) 
in the following respects. 
(i) The spectral weight with a large peak at $\omega/t=2$ in the 
$U/t\rightarrow 0$ limit [see Fig.~\ref{fig1}(d)] is partially 
transferred to a broad higher-energy region corresponding to the 
``upper Hubbard band'' with increasing $U/t$, which is observed in 
both the VCA and CDIA.  
(ii) With increasing $U$, the remaining spectral weight at 
$\omega/t\simeq 2$ shifts to higher energies in the VCA [see Figs.~\ref{fig3}(c) and \ref{fig3}(d)], while in the CDIA, it shifts rapidly to lower energies and 
simultaneously loses its weight [see Figs.~\ref{fig5}(c) and \ref{fig5}(d)].  
(iii) We thus have a large spectral weight at low energies ($\omega/t\alt 1$) 
in the CDIA, which is rather small in the VCA.  The spectral weight characteristic 
of the massless Dirac SM dispersions can, however, be seen in the vicinity 
of the Fermi level in both the VCA and CDIA spectra.  
(iv) More quantitatively, a kink appears in the lowest-energy 
region of the DOS in the VCA (see the inset of the lower panels of 
Fig.~\ref{fig3}), which shifts toward the Fermi level with increasing $U$.  
The DOS curve becomes steeper near the Fermi level (or 
the ${\bm k}$-linear dispersion becomes flatter at the Dirac point), 
renormalizing the Fermi velocity but keeping the electrons massless.  
No quasiparticle peak appears.  At a critical $U$ value, the kink disappears 
and simultaneously the gap begins to open gradually.  In the CDIA, similar 
but stronger effects can be seen with increasing $U/t$ in the lowest-energy 
region of the DOS, until the gap opens discontinuously at $U_c$ 
[see Figs.~\ref{fig5}(c) and \ref{fig5}(d)].  These low-energy behaviors in the CDIA 
are consistent with the results of the single-site DMFT for the honeycomb 
Hubbard model\cite{MK09,J09} and are expected to be realistic in the 
honeycomb and $\pi$-flux Hubbard models where the electron-number fluctuates.  

\begin{figure}[t]
\begin{center}
\includegraphics[width=\columnwidth]{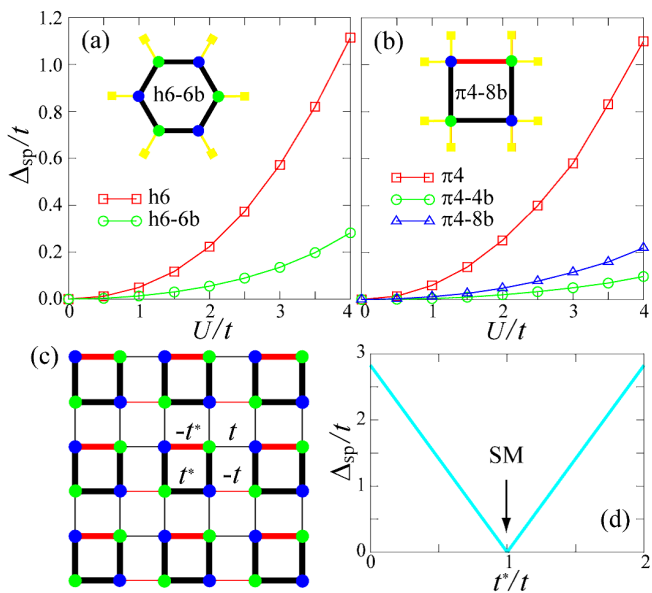}
\caption{(Color online) 
Calculated single-particle gap $\Delta_{\mathrm{sp}}$ in the PM state of 
the (a) honeycomb and (b) $\pi$-flux Hubbard models.  For the honeycomb 
lattice, we use the hexagonal 6-site cluster (h6) and 6-site 6-bath cluster 
(h6-6b).  For the $\pi$-flux lattice, we use the square 4-site cluster 
($\pi$4), 4-site 4-bath cluster ($\pi$4-4b), and 4-site 8-bath cluster 
($\pi$4-8b).  
(c) $\pi$-flux lattice with renormalized hopping parameter $t^*$ 
violating the original translational symmetry and (d) its noninteracting 
single-particle gap as a function of $t^*/t$.  
}\label{fig6}
\end{center}
\end{figure}

\subsection{Cluster geometry dependence}

Finally, let us discuss the cluster geometry dependence of the 
single-particle gap $\Delta_{\mathrm{sp}}$ in the PM phase.  
In Figs.~\ref{fig6}(a) and \ref{fig6}(b), we show the results of the 6-site 6-bath 
system for the honeycomb Hubbard model and of the 4-site 4-bath and 
4-site 8-bath systems for the $\pi$-flux Hubbard model.  
In the honeycomb lattice, we find that even if we add the bath sites, 
the gap $\Delta_{\mathrm{sp}}$ opens at any infinitesimal $U$ value when 
we use the 6-site hexagonal ring cluster as the reference 
system.\cite{HL12,SO12,HS13} 
In the $\pi$-flux Hubbard model, we also find that the gap $\Delta_{\mathrm{sp}}$ 
opens at any infinitesimal $U$ value when we use the 4-site square cluster 
as the reference system. 
Therefore, even though we use two bath sites per correlation site, 
the gap opens at infinitesimal $U$ values, which does not agree with the 
argument in Ref.~\citen{HS13} that at least two bath sites per correlation 
site are necessary to discuss the MIT in the honeycomb lattice.  

Rather, our results agree with the statement in Ref.~\citen{LW13} that the 
opening of the gap at infinitesimal $U/t$ values is not caused by the 
bath degrees of freedom but by the cluster geometry, which violates the 
original translational symmetry of the lattice.  We show the latter case 
in Fig.~\ref{fig6}(c), where the original translational symmetry of the 
$\pi$-flux lattice is violated by the renormalization of the hopping parameter $t^*$ 
by the interaction only within the cluster, leading to 
$t^*\ne t$.  Then, as shown in Fig.~\ref{fig6}(d), the noninteracting 
band with $t^*$ and $t$ has a finite single-particle gap unless $t^*=t$.  
A similar discussion has been given for the honeycomb lattice,\cite{WRLL12,LRTR14} 
where the 6-site hexagonal clusters with the renormalized hopping parameter 
$t^*$ are connected with the bare hopping parameter $t$.  
A ``plaquette insulator'' state is thus realized at $t^*\ne t$ in the 
noninteracting limit.  
This is the reason why the single-particle gap opens at infinitesimal 
$U/t$ values.  
However, we here point out that it is always possible to make an appropriate 
choice of clusters that maintains the Dirac zero-gap situation even 
though it violates the original translational symmetry, examples of 
which are shown in Figs.~\ref{fig2}(d)-\ref{fig2}(f) where the gap does not open 
at small values of $U$.  Thus, the statement in Ref.~\citen{LW13} is 
too strict.  
Careful choice of the clusters in the quantum cluster methods such as 
CDMFT, the VCA, and the CDIA enables one to discuss the MIT of Dirac fermion 
systems without spurious opening of the gap.  

\section{Summary}

We have made a comparative study on the MIT of Dirac 
electrons in the honeycomb and $\pi$-flux Hubbard models 
using the VCA and CDIA, where we have calculated the single-particle 
gap and staggered magnetization as functions of the interaction 
strength $U$.  
We have paid particular attention to the choice of the cluster 
geometry and the inclusion of the bath sites.  We have thus confirmed 
that the spurious single-particle gap that opens at infinitesimal 
$U$ values is not caused by the bath degrees of freedom but 
rather by the cluster geometry.  
We have shown that with increasing $U$, the first-order MIT to the nonmagnetic MI phase occurs in the presence of electron-number fluctuation.  
However, the AFMI phase always preempts this MIT, at least in the 
present models, and therefore the SL phase previously suggested 
to emerge between the Dirac SM and AFMI phases is absent in 
these models.  

Our results imply that, if the AF ordering can be suppressed by, 
for example, to the effect of spin frustrations in the triangular and 
related lattices, one may expect that the MI phase without AF 
orders will preempt the AFMI phase, resulting in the emergence 
of the SL phase in the intermediate coupling region, as was 
pointed out recently in Ref.~\citen{RLRT14} for the triangular 
$\pi$-flux Hubbard model.  

\bigskip
\acknowledgments{
We thank K. Seki for enlightening discussions.  
T.~K. acknowledges support from a JSPS Research Fellowship for 
Young Scientists.  
This work was supported in part by KAKENHI Grant No.~26400349 
from JSPS of Japan.
}

\end{document}